\documentstyle[preprint,aps]{revtex}
\begin{document}
\tighten

\def\beq{\begin{equation}}
\def\eeq{\end{equation}}
\def\beqy{\begin{eqnarray}}
\def\eeqy{\end{eqnarray}}
\def\pr#1#2#3{ {\sl Phys. Rev.\/} {\bf#1}, #3 (#2)}
\def\prl#1#2#3{ {\sl Phys. Rev. Lett.\/} {\bf#1}, #3 (#2)}
\def\np#1#2#3{ {\sl Nucl. Phys.\/} {\bf#1}, #3 (#2)}
\def\cmp#1#2#3{ {\sl Comm. Math. Phys.\/} {\bf#1}, #3 (#2)}
\def\pl#1#2#3{ {\sl Phys. Lett.\/} {\bf#1}, #3 (#2)}
\def\apj#1#2#3{ {\sl Ap. J.\/} {\bf#1}, #3 (#2)}
\def\aop#1#2#3{ {\sl Ann. Phy.\/} {\bf#1}, #3 (#2)}
\def\nc#1#2#3{ {\sl Nuovo Cimento }{\bf#1}, #3 (#2)}
\def\cjp#1#2#3{ {\sl Can. J. Phys. }{\bf#1}, #3 (#2)}
\def\zp#1#2#3{ {\sl Z. Phys. }{\bf#1}, #3 (#2)}
\def\pbi{\vec{p}_i}
\def\pbj{\vec{p}_j}
\def\yc{{\cal Y}}
\def\tp0{^3P_0}
\def\nii{\noindent}
\def\pbb{{\bf p}}
\def\Pbb{{\bf P}}
\def\qbb{{\bf q}}
\def\kbb{{\bf k}}
\def\Kbb{{\bf K}}
\def\yc{{\cal Y}}
\def\he{\hat \ell}
\def\hs{\hat S}
\def\hl{\hat L}
\def\hj{\hat J}
\def\rmb#1{{\bf #1}}
\def\lpmb#1{\mbox{\boldmath $#1$}}
\def\half{{\textstyle{1\over2}}}
\def\thalf{{\textstyle{3\over2}}}
\def\fhalf{{\textstyle{5\over2}}}
\def\shalf{{\textstyle{7\over2}}}
\def\nhalf{{\textstyle{9\over2}}}
\def\lhalf{{\textstyle{11\over2}}}
\def\tthalf{{\textstyle{13\over2}}}
\def\fthalf{{\textstyle{15\over2}}}
\def\nn{\nonumber}
\def\>{\rangle}
\def\<{\langle}

\title{New Baryons in the $\Delta \eta$ and $\Delta \omega$ Channels}
\author{Simon Capstick}
\address{Supercomputer Computations Research Institute
  and Department of Physics,\\
  Florida State University,
  Tallahassee, FL 32306}
\author{W. Roberts}
\address{Department of Physics, Old Dominion University\\
Norfolk, VA 23529 USA \\ and \\
Thomas Jefferson National Accelerator Facility\\
12000 Jefferson Avenue, Newport News, VA 23606, USA}
\date{\today}
\maketitle

\begin{flushright}
FSU-SCRI-97-95

JLAB-TH-97-38
\end{flushright}

\begin{abstract}
The decays of excited nonstrange baryons into the final states $\Delta
\eta$ and $\Delta\omega$ are examined in a relativized quark pair
creation model. The wavefunctions and parameters of the model are
fixed by previous calculations of $N \pi$ and $N \pi\pi$, {\it etc.}, decays
through various quasi-two body channels including $N\eta$ and
$N\omega$. Our results show that the combination of thresholds just
below the region of interest and the isospin selectivity of these
channels should allow the discovery of several new baryons in such
experiments.
\end{abstract}  
\newpage 
\def\slash#1{#1 \hskip -0.5em / }  

\section{Introduction}

Quark models of baryon structure based on three effective quark
degrees of freedom predict the existence of more states than have
previously been seen in analyses of $N\pi$ elastic scattering. In
particular, there are approximately nine `missing' states predicted by
quark potential models to lie in the first band of positive-parity
excited states (which we define as states whose wavefunctions are
predominantly $N=2$ band when expanded in a harmonic oscillator
basis). One of these states, a second $N\thalf^+$ ($P_{13}$) resonance,
may have been discovered in the coupled channel analysis of Manley and
Saleski~\cite{MANSA}. There remain six missing nucleon and two missing
$\Delta$ states with model masses between approximately 1850 and 2050
MeV. There are also many undiscovered states predicted by these models
which have wavefunctions which lie predominantly in the $N=3$ and
higher bands, the lightest of which are predicted to have negative
parity~\cite{spectrum,CI}.

Models of this kind, when combined with a model of the strong decays
of baryon states~\cite{decays}, yield a simple explanation for the
absence of the missing states~\cite{KI,SS,scwr,BIL} in analyses of
$N\pi$ elastic scattering--they simply have weak $N\pi$ couplings and
so contribute little to $N\pi$ scattering amplitudes in their partial
waves. A simple solution is to produce these states
electromagnetically with real photons or by electron scattering, and
then look for their decays to final states other than
$N\pi$~\cite{NI}. As part of the $N^*$ program in Hall B at the Thomas
Jefferson National Accelerator Facility (TJNAF), an
experiment~\cite{93033} will study $\gamma p \to p\pi^+\pi^-$, with
analysis focusing on $\gamma p\to \Delta^{++}\pi^-$, $\gamma p\to
\Delta^{0}\pi^+$ and $\gamma p\to p\rho^0$.  Previous theoretical
work~\cite{KI,SS,CR2,BIL} has shown that several of the missing states
and many of the undiscovered states have sizeable couplings to these
channels. Other experiments will focus on the decays of such states to
$N\eta$~\cite{89039}, $N\eta$ and $N\eta^\prime$~\cite{91008}, and
$N\omega$~\cite{91024}.  These channels offer the advantage of being
isospin selective, in that only $I=\half$ $N^*$ resonances (as opposed
to $I=\thalf$ $\Delta^*$ resonances) can couple to these final
states. Given the predicted near degeneracy of broad states in several
partial waves in this region, this isospin selectivity should simplify
what is likely to be a difficult analysis to extract information about
these states.

Detection and analysis of the final states $\Delta(1232)\eta$ and
$\Delta\omega$ in electromagnetic production from protons at TJNAF
will be complicated by the increased particle multiplicity, and by the
presence of an additional neutral particle in the final state
resulting from the decays $\Delta^+\to p\pi^0,\ n\pi^+$. For these
experiments, it may be better to produce these final states from the
neutron in the deuteron~\cite{JimNap}, reconstruct the $\Delta^0\to
p\pi^+$ charged particle decay, and use missing mass to identify the
$\eta$ or $\omega$. At the AGS, the properties of the Crystal Barrel
detector make it ideal for examining the process $\pi^- p\to
n\pi^0\eta$. The final state in $\pi^+p\to p\pi^+\eta$ is more difficult
to detect but can, in principle, also be seen using the Crystal Barrel
\cite{nefkens}.  Despite these detection difficulties, these channels
also have the advantage of being isospin selective, and can in
principle isolate the two missing $\Delta$ resonances and higher lying
$\Delta$ states if they are present and are produced.

Another advantage of a $\Delta\eta$ experiment is that the threshold
for this reaction lies just below the mass region where these states
are predicted.  The nominal $\Delta\eta$ threshold is at 1780 MeV; as
we integrate over the lineshape of the $\Delta$, the effective
threshold is at $m_N+m_{\pi}+m_\eta\simeq 1630$ MeV. This is to be
compared to mass predictions~\cite{spectrum,CI} for the lightest
missing states of around 1800-1850 MeV $[\Delta\half^+]_1$ and around
1950-2000 MeV for $[\Delta\thalf^+]_4$ (which would be a first
$P_{31}$ and a fourth $P_{33}$ state in $N\pi$, respectively). It is
generally true (in decay models and in experiment) that once the
energy available for a decay increases beyond the region where the
phase space has initially become appreciable, the decay amplitudes
tend to decrease rapidly as the three momentum available to the final
particles increases and the wavefunction overlaps diminish.

Here we provide predictions for the decay amplitudes into the final
states $\Delta\eta$ and $\Delta\omega$ of all states (missing and seen
in $N\pi$) with wavefunctions predominantly in the $N=1$ and $N=2$
bands, and also for several low-lying states in higher bands, using
the relativized model of baryon decays based on the $^3P_0$ pair
creation model of Refs.~\cite{scwr} and~\cite{CR2}. Decays into the
$\Delta\eta$ channel have been previously considered by Bijker,
Iachello and Leviatan in Ref.~\cite{BIL}, within an algebraic model of
the spectrum and wavefunctions, using pointlike emitted mesons.

In the present calculation model parameters are taken from our
previous work and not adjusted. Wavefunctions are taken from the
relativized model of Ref.~\cite{CI}, which describes all of the states
considered here in a consistent picture. In order to be in accord with
the Particle Data Group (PDG)~\cite{PDG} conventional definitions of
decay widths, we have determined the decay momentum using the central
value of the PDG quoted mass for resonances seen in $N\pi$, and the
predicted mass from Ref.~\cite{CI} for missing and undiscovered
states. We have also integrated over the lineshape of the final
$\Delta$ baryon, with the final phase space as prescribed in the meson
decay calculation of Ref.~\cite{KokI}; for details of this procedure
see Eq.~(8) of Ref.~\cite{CR2} (note that we do not integrate over the
narrow [8 MeV width] $\omega$ lineshape). As a consequence there are
states below the nominal thresholds which have non-zero decay
amplitudes.

In keeping with the convention of Ref.~\cite{CR2}, the phases of the
amplitudes are determined as follows. We quote the product
$A^{X\dag}_{\Delta M}A^X_{N\pi}/|A^X_{N\pi}|$ of the predicted decay
amplitude for $X\to \Delta M$ (where $M$ is either $\eta$ or $\omega$)
and the phase of the decay amplitude for $X\to N\pi$, the latter being
unobservable in $N\pi$ elastic scattering (note factors of $+i$,
conventionally suppressed in quoting amplitudes for decays of negative
parity baryons to $NM$ or $N\gamma$, where $M$ has negative parity, do
not affect this product). This eliminates problems with (unphysical)
sign conventions for wavefunctions, and the relative signs of these
products are then predictions for the (physically significant)
relative phases of the contributions of states $X$ in the process
$N\pi\to X\to \Delta M$. Since the missing and undiscovered states may
have small $N\pi$ couplings it may be useful to find the relative
signs of the contributions of states $X$ in the process $N\gamma\to
X\to \Delta\eta$. As the photocouplings of Ref.~\cite{Cpc} are also
quoted inclusive of the $N\pi$ sign,
$A^{X\dag}_{N\gamma}A^X_{N\pi}/|A^X_{N\pi}|$, then simply multiplying
the quoted photocouplings by the amplitudes quoted here will yield the
relative phases of the contributions of states $X$ in $N\gamma\to X\to
\Delta M$.

We note that we have chosen the meson wave flavor functions as
\begin{eqnarray}
\eta&=&\frac{1}{\sqrt{2}}\left[\frac{1}{\sqrt{2}}\left(u\overline{u}+
d\overline{d}\right)-s\overline{s}\right]\nonumber\\
\eta^\prime&=&\frac{1}{\sqrt{2}}\left[\frac{1}{\sqrt{2}}\left(u\overline{u}+
d\overline{d}\right)+s\overline{s}\right]\nonumber\\
\omega&=&\frac{1}{\sqrt{2}}\left(u\overline{u}+d\overline{d}\right),
\end{eqnarray}
{\it i.e.} we allow for ideal mixing between $\omega$ and
$\phi=s\overline{s}$, and an $\eta$--$\eta^\prime$ mixing angle of
$\theta_P=-9.7\,^{\rm o}$.

\section{Results and Discussion}

Our results are given in Tables~\ref{DNle2Deta} to~\ref{DNge3Domega},
where we list the model state, its assignment (if any) to a resonance
from the analyses, and its decay amplitudes into the $\Delta\eta$ and
$\Delta\omega$ channels. The predictions for the $N\pi$ decay
amplitudes for each state~\cite{scwr} and values for these amplitudes
extracted from the PDG~\cite{PDG} are also included for ease of
identification of missing resonances. All theoretical amplitudes are
given with upper and lower limits, along with the central value, in
order to convey the uncertainty in our results due to the uncertainty
in the resonance's mass.  These correspond to our predictions for the
amplitudes for a resonance whose mass is set to the upper and lower
limits, and to the central value, of the experimentally determined
mass. For states as yet unseen in the analyses of the data, we have
adopted a `standard' uncertainty in the mass of 150 MeV and used the
model predictions for the state's mass for the central value. If a
state below the effective threshold has been omitted from a table it
is because our predictions for all of its amplitudes are zero.

For completeness we have also calculated decays to the
$\Delta\eta^\prime$ channel, and find that all of the amplitudes for
the states considered here are small. This is primarily due to the
high effective (nominal) threshold of roughly 2040 (2190) MeV. We do not
record these amplitudes here.

\subsection{$\Delta\eta$ decays}

The results for this channel are shown in Tables \ref{DNle2Deta} and
\ref{DNge3Deta}. Amplitudes for the lighter states are predictably
small due to the effective (nominal) threshold of 1638 (1780) MeV,
with some notable exceptions. In Table \ref{DNle2Deta}, the
largest amplitudes are those of the $\Delta(2000)F_{35}$ state, which
is a two-star state~\cite{PDG}. This channel thus offers a very good
opportunity for confirmation of this state. In addition, it should be
possible to detect the missing fourth $P_{33}$ state
$[\Delta\frac{3}{2}^+]_4(1985)$, and to confirm the first $P_{31}$
state seen in the multichannel analysis of Ref.~\cite{MANSA}, as their
couplings to this channel are appreciable.

In table \ref{DNge3Deta}, we see that it may be possible to confirm
the one-star states $\Delta(1940)D_{35}$ and $\Delta(2390)F_{37}$ in a
$\Delta\eta$ experiment. Our results predict that the model states
$[\Delta\frac{3}{2}^-]_3(2145)$, $[\Delta\frac{5}{2}^-]_2(2165)$,
$[\Delta\frac{7}{2}^-]_1(2230)$, $[\Delta\frac{7}{2}^-]_2(2295)$ and
$[\Delta\frac{9}{2}^+]_2(2505)$ offer the best opportunities for
discovery in this channel.

\subsection{$\Delta\omega$ decays}

The high effective (nominal) threshold of approximately 1860 (2010) MeV
precludes sizeable couplings of states with wavefunctions
predominantly below the $N=3$ band to the $\Delta\omega$ channel (see
Table~\ref{DNle2Domega}), although there are some states
with amplitudes which grow rapidly away from the threshold, and so
will couple if the actual mass is larger than the nominal mass.  Key
examples are the missing $[\Delta\frac{3}{2}^+]_4(1985)$ and the
two-star $\Delta(2000)F_{35}$. However, there are several of the more
highly excited states considered here in Table~\ref{DNge3Domega} with
appreciable couplings to this channel. It may be possible to confirm
the weak states $\Delta(2150)S_{31}$ (one star), $\Delta(2400)G_{39}$
(two stars), and $\Delta(2390)F_{37}$ (one star) in this channel, or
perhaps even the very highly excited states $\Delta(2750)I_{3\,13}$
and $\Delta(2950)K_{3\,15}$ (both two-star states). Our results 
predict that a $\Delta\omega$ experiment may also be able to discover
several predicted states, the most interesting of which are
$\Delta[\frac{3}{2}^-]_3(2145)$, the two states
$[\Delta\frac{5}{2}^-]_2(2165)$ and $[\Delta\frac{5}{2}^-]_3(2265)$,
the two states $[\Delta\frac{7}{2}^-]_1(2230)$and
$[\Delta\frac{7}{2}^-]_2(2295)$, and $[\Delta\frac{9}{2}^+]_2(2505)$.

\subsection{Conclusions}

Our results show that it should be possible to discover in a
$\Delta\eta$ experiment one of the two low-lying states missing from
previous $N\pi$ analyses, and to confirm the possible discovery of
another. The fact that these are the first states in their partial waves to
couple to this channel will make their extraction from an analysis
less complicated than in final states with lower thresholds. There are
also two weakly established states in the 1900--2000 MeV range which
may be confirmed in such an experiment, and several higher mass
predicted states which couple appreciably to this channel.

Amplitudes for all states to couple to the $\Delta\eta^\prime$ channel
are small, largely due to the high threshold. This is not true of the
$\Delta\omega$ channel. Our results show that above 2150 MeV such an
experiment may allow confirmation of several weakly established states
and the discovery of a substantial number of predicted states.

Reconstruction of the $\Delta\eta$ and $\Delta\omega$ final states
will be difficult due to the final state particle multiplicity and the
presence of neutral particles in electromagnetic production from the
proton. However it may be that the extraction of information about
these important new baryon states from an analysis of the results of
such an experiment is considerably less complicated than using
channels with fewer final state particles. This indicates that it is
worthwhile to consider developing such experiments.

\section{Acknowlegements}
The importance of these channels was pointed out to us by Professors
D. Mark Manley and B. Nefkens. This work was supported in part by the
the Florida State University Supercomputer Computations Research
Institute which is partially funded by the Department of Energy
through Contract DE-FC05-85ER250000 (SC); the U.S.  Department of
Energy under Contract No. DE-FG05-86ER40273 (SC); the National Science
Foundation through Grant No.\ PHY-9457892 (WR); the U.S.  Department
of Energy under Contract No.\ DE-AC05-84ER40150 (WR); and by the
U.S. Department of Energy under Contract No.\ DE-FG02-97ER41028 (WR).

\begin{table}

\caption{Results for $\Delta$ states in the $N=1$ and $N=2$ bands in
the $\Delta\eta$ channel. $N\pi$
amplitudes from Ref.~\protect{\cite{scwr}} are included to explain our
assignments of the model states to resonances. Notation
for model states is $[J^P]_n({\rm mass[MeV]})$, where $J^P$ is the
spin/parity of the state and $n$ its principal quantum number. The
first row gives our model results, while the second row lists the
$N\pi$ amplitudes from the partial-wave analyses, as well as the
Particle Data Group (PDG) name for the state, its $N\pi$ partial wave,
and its PDG star rating. Light states with zero amplitudes are omitted
from the table.}

\label{DNle2Deta}
\begin{tabular}{@{}l@{\hspace{6pt}}r@{\hspace{6pt}}r@{\hspace{6pt}}r@{\hspace{6
pt}}r@{\hspace{6pt}}r@{\hspace{6pt}}r@{\hspace{6pt}}r@{}}
\multicolumn{1}{c} {model state}
& \multicolumn{1}{r}{$N\pi$}
& \multicolumn{1}{r}{$\Delta\eta$}
& \multicolumn{1}{r}{$\Delta\eta$}
& \multicolumn{1}{r}{$\sqrt{\Gamma^{\rm tot}_{\Delta\eta}}$}
\\ 
\multicolumn{1}{c} {$N\pi$ state/rating} & & & & \\ 
\tableline
& & $s$ & $d$ & \\
\cline{3-4}

$[\Delta\textstyle{{3\over2}}^-]_1(1620)$ & 4.9 $\pm $ 0.7 & 1.1 $^{+
 3.2}_{- 1.1}$ & 0.0 $^{+ 0.3}_{- 0.0}$ & 1.1 $^{+ 3.2}_{- 1.1}$\\%[-9pt]

$\Delta(1700)D_{33}$**** & $6.5\pm 2.0$ \\

& & $p$ & & \\
\cline{3-3}

$[\Delta\textstyle{{1\over2}}^+]_1(1835)$ & 3.9 $^{+ 0.4}_{- 0.7}$ & 3.2
    $^{+ 4.1}_{- 3.1}$ & & 3.2 $^{+ 4.1}_{- 3.1}$\\%[-9pt]

$\Delta(1740)P_{31}$$^{\rm a}$ & $4.9\pm 1.3$ \\

$[\Delta\textstyle{{1\over2}}^+]_2(1875)$ & 9.4 $\pm $ 0.4 & -2.9 $\pm
 $ 0.7 & & 2.9 $\pm $ 0.7 \\%[-9pt]

$\Delta(1910)P_{31}$**** & $6.6\pm 1.6$ \\

& & $p$ & $f$ & \\
\cline{3-4}

$[\Delta\textstyle{{3\over2}}^+]_2(1795)$ & 8.7 $\pm $ 0.2 & 0.0 $^{+
0.3}_{- 0.0}$ & 0.0 $\pm $ 0.0 & 0.0 $^{+ 0.3}_{- 0.0}$\\%[-9pt]

$\Delta(1600)P_{33}$*** & $7.6\pm 2.3$ \\

$[\Delta\textstyle{{3\over2}}^+]_3(1915)$ & 4.2 $\pm $ 0.3 & -3.3
$\pm $ 0.9 & 0.7 $\pm $ 0.4 & 3.4 $\pm $ 0.9 \\%[-9pt]

$\Delta(1920)P_{33}$*** & $7.7\pm 2.3$ \\

$[\Delta\textstyle{{3\over2}}^+]_4(1985)$ & 3.3 $^{+ 0.8}_{- 1.1}$ &
-4.2 $^{+ 2.4}_{- 1.7}$ & - 0.7 $^{+ 0.6}_{- 1.2}$ & 4.3 $^{+ 1.9}_{-
2.5}$ \\

& & $p$ & $f$ & \\
\cline{3-4}

$[\Delta\textstyle{{5\over2}}^+]_1(1910)$ & 3.4 $\pm $ 0.2 & -0.5 $\pm
$ 0.1 & 0.6 $\pm $ 0.3 & 0.8 $\pm $ 0.3 \\%[-9pt]

$\Delta(1750)F_{35}$$^{\rm b}$ & $2.0\pm 0.8$ \\%[-9pt]
$\Delta(1905)F_{35}$**** & $5.5\pm 2.7$ \\

$[\Delta\textstyle{{5\over2}}^+]_2(1990)$ & 1.2 $\pm $ 0.4 & -7.0 $^{+
 5.1}_{- 2.9}$ & 0.3 $^{+ 0.8}_{- 0.3}$ & 7.0 $^{+ 2.9}_{- 5.1}$ \\%[-9pt]

$\Delta(2000)F_{35}$** & $5.3\pm 2.3$\\

& & $f$ & $h$ & \\ 
\cline{3-4} 

$[\Delta\textstyle{{7\over2}}^+]_1(1940)]$ & 7.1 $\pm $ 0.1 & 0.9 $\pm
$ 0.1 & 0.0 $\pm $ 0.0 & 0.9 $\pm $ 0.1 & \\%[-9pt]

$\Delta(1950)F_{37}$**** & $9.8\pm 2.7$ \\
\tableline
\noalign{a\ \ First $P_{31}$ state found in Ref.~\protect{\cite{MANSA}}.}
\noalign{b\ \ Ref.~\protect{\cite{MANSA}} finds two $F_{35}$ states; this one 
and 
$\Delta(1905)F_{35}$.}
\end{tabular}
\end{table}
\begin{table}

\caption{Results in the $\Delta\eta$ channel for the lightest few
negative-parity $\Delta$ resonances of each $J$ in the N=3 band, and
for the lightest few $\Delta$ resonances for $J^P$ values which first
appear in the N=4, 5 and 6 bands. Notation as in
Table~\protect{\ref{DNle2Deta}}.}

\label{DNge3Deta}

\begin{tabular}{@{}l@{\hspace{6pt}}r@{\hspace{6pt}}r@{\hspace{6pt}}r@{\hspace{6
pt}}r@{\hspace{6pt}}r@{\hspace{6pt}}r@{\hspace{6pt}}r@{}}
\multicolumn{1}{c} {model state}
& \multicolumn{1}{r}{$N\pi$}
& \multicolumn{1}{r}{$\Delta\eta$}
& \multicolumn{1}{r}{$\Delta\eta$}
& \multicolumn{1}{r}{$\sqrt{\Gamma^{\rm tot}_{\Delta\eta}}$}
\\ 
\multicolumn{1}{c} {$N\pi$ state/rating} & & & & \\ 
\tableline
& & $d$ \\
\cline{3-3}

$[\Delta\half^-]_2(2035)$ & 1.2 $\pm $ 0.2 & 1.8 $^{+ 0.9}_{- 0.7}$ & &
1.8 $^{+ 0.9}_{- 0.7}$ \\%[-9pt]

$\Delta\half^-(1900)S_{31}$*** & $4.1\pm 2.2$ \\

$[\Delta\half^-]_3(2140)$ & 3.1 $^{+ 0.4}_{- 1.1}$ & -2.4 $^{+ 1.0}_{-
.6}$ & & 2.4 $^{+ 0.6}_{- 1.0}$ \\%[-9pt]

$\Delta\half^-(2150)S_{31}$* & $4.0\pm 1.5$\\

& & $s$ & $d$ \\
\cline{3-4}

$[\Delta\thalf^-]_2(2080)$ & 2.1 $\pm $ 0.1 & 2.4 $\pm $ 0.6 & 1.9 $^{+
1.9}_{- 1.6}$ & 3.1 $^{+ 1.8}_{- 1.3}$ \\%[-9pt]

$\Delta\thalf^-(1940)D_{33}$* & $3.2\pm 1.4$\\

$[\Delta\thalf^-]_3(2145)$ &2.2 $^{+ 0.1}_{- 0.3}$ & -1.9 $^{+ 0.4}_{-
1.2}$ & 3.3 $^{+ 0.9}_{- 1.5}$ & 3.8 $\pm $ 1.4 \\

& & $d$ & $g$ \\
\cline{3-4}

$[\Delta\fhalf^-]_1(2155)$ & 5.2 $\pm $ 0.0 & 1.1 $\pm $ 0.3 & -0.1 $\pm
$ 0.0 & 1.1 $\pm $ 0.3 \\%[-9pt]

$\Delta\fhalf^-(1930)D_{35}$*** & $5.0\pm 2.3$\\

$[\Delta\fhalf^-]_2(2165)$ & 0.6 $\pm $ 0.1 & 3.7 $^{+ 0.9}_{- 1.6}$ &
1.3 $^{+ 1.2}_{- 0.9}$ & 3.9 $^{+ 1.3}_{- 1.8}$ \\

$[\Delta\fhalf^-]_3(2265)$ & 2.4 $\pm $ 0.4 & -2.7 $\pm $ 0.2 & 1.2
$\pm $ 0.4 & 2.9 $\pm $ 0.4 \\

$[\Delta\fhalf^-]_4(2325)$ & 0.1 $\pm $ 0.0 & -2.4 $^{+ 0.4}_{- 0.1}$ &
 1.1 $^{+ 0.7}_{- 0.5}$ & 2.6 $^{+ 0.5}_{- 0.6}$ \\

$[\Delta\shalf^-]_1(2230)$ & 2.1 $\pm $ 0.6 & 3.8 $^{+ 0.6}_{- 1.5}$ &
1.2 $^{+ 1.0}_{- 0.8}$ & 4.0 $^{+ 0.9}_{- 1.7}$ \\

$[\Delta\shalf^-]_2(2295)$ & 1.8 $\pm $ 0.4 & -4.0 $^{+ 1.0}_{- 0.3}$ &
    1.5 $^{+ 0.9}_{- 0.8}$ & 4.2 $^{+ 0.6}_{- 1.1}$ \\

& & $g$ & $i$ \\
\cline{3-4}

$[\Delta\nhalf^-]_1(2295)$ & 4.8 $\pm $ 1.3 & 2.2 $^{+ 2.1}_{- 1.2}$
& 0.0 $\pm $ 0.0 & 2.2 $^{+ 2.1}_{- 1.2}$ \\%[-9pt]

$\Delta\nhalf^-(2400)G_{39}$** & $4.1\pm 2.1$\\

& & $f$ & $h$ \\
\cline{3-4}

$[\Delta\shalf^+]_2(2370)$ & 1.5 $^{+ 0.6}_{- 0.9}$ & 2.7 $^{+ 0.4}_{-
0.6}$ & 0.0 $\pm $ 0.0 & 2.7 $^{+ 0.4}_{- 0.6}$ \\%[-9pt]

$\Delta\shalf^+(2390)F_{37}$* & $4.9\pm 2.0$\\

$[\Delta\shalf^+]_3(2460)$ & 1.1 $^{+ 0.0}_{- 0.1}$ & -1.6 $\pm $ 0.4 &
 1.0 $^{+ 0.9}_{- 0.5}$ & 1.9 $^{+ 0.9}_{- 0.6}$ \\

$[\Delta\nhalf^+]_1(2420)$ & 1.2 $\pm $ 0.4 & -0.2 $\pm $ 0.1 & 0.7 $^{+
 0.7}_{- 0.4}$ & 0.7 $^{+ 0.7}_{- 0.4}$ \\%[-9pt]

$\Delta\nhalf^+(2300)H_{39}$** & $5.1\pm 2.2$\\

$[\Delta\nhalf^+]_2(2505)$ & 0.4 $\pm $ 0.1 & -3.3 $\pm $ 0.7 & 0.3 $^{+
 0.3}_{- 0.1}$ & 3.3 $\pm $ 0.8 \\

& & $h$ & $j$ \\
\cline{3-4}

$[\Delta\lhalf^+]_1(2450)$ & 2.9 $\pm $ 0.7 & 1.0 $^{+ 0.7}_{- 0.4}$ & 0.0
$\pm $ 0.0 & 1.0 $^{+ 0.7}_{- 0.4}$ \\%[-9pt]

$\Delta\lhalf^+(2420)H_{3\,\,11}$**** & $6.7\pm 2.8$\\

$[\Delta\tthalf^+]_1(2880)$ & 0.8 $\pm $ 0.2 & 0.0 $\pm $ 0.0 & 1.3 $^{+
 0.8}_{- 0.6}$ & 1.3 $^{+ 0.8}_{- 0.6}$ \\

$[\Delta\tthalf^+]_2(2955)$ & 0.2 $\pm $ 0.1 & -2.1 $^{+ 0.4}_{- 0.2}$ &
0.3 $\pm $ 0.1 & 2.1 $^{+ 0.2}_{- 0.4}$ \\

& & $i$ & $k$ \\
\cline{3-4}

$[\Delta\tthalf^-]_1(2750)$ & 2.2 $\pm $ 0.4 & 1.6 $^{+ 0.7}_{- 0.5}$ &
0.0 $\pm $ 0.0 & 1.6 $^{+ 0.7}_{- 0.5}$ \\%[-9pt]

$\Delta\tthalf^-(2750)I_{3\,\,13}$** & $3.7\pm 1.5$\\

& & $j$ & $l$ \\
\cline{3-4}

$[\Delta\fthalf^+]_1(2920)$ & 1.6 $\pm $ 0.3 & 1.4 $\pm $ 0.5 & 0.0 $\pm
$ 0.0 & 1.4 $\pm $ 0.5 \\%[-9pt] 

$\Delta\fthalf^+(2950)K_{3\,\,15}$** & $3.6\pm 1.5$\\

$[\Delta\fthalf^+]_2(3085)$ & 0.4 $\pm $ 0.1 & 0.4 $\pm $ 0.2 & 0.0 $\pm $
0.0 & 0.4 $\pm $ 0.2 \\

\end{tabular}
\end{table}
\begin{table}

\caption{Results for $\Delta$ states in the $N=1$ and $N=2$ bands in
the $\Delta\omega$ channel. Notation as in Table~\protect{\ref{DNle2Deta}}.}

\label{DNle2Domega}

\begin{tabular}{@{}l@{\hspace{6pt}}r@{\hspace{6pt}}r@{\hspace{6pt}}r@{\hspace{6
pt}}r@{\hspace{6pt}}r@{\hspace{6pt}}r@{\hspace{6pt}}r@{}}
\multicolumn{1}{c} {model state}
& \multicolumn{1}{r}{$\Delta\omega$}
& \multicolumn{1}{r}{$\Delta\omega$}
& \multicolumn{1}{r}{$\Delta\omega$}
& \multicolumn{1}{r}{$\Delta\omega$}
& \multicolumn{1}{r}{$\Delta\omega$}
& \multicolumn{1}{r}{$\Delta\omega$}
& \multicolumn{1}{r}{$\sqrt{\Gamma^{\rm tot}_{\Delta\omega}}$}
\\ 
\multicolumn{1}{c} {$N\pi$ state/rating} & & & & & & &\\ 
\tableline

& $p_\half$ & $p_\thalf$ & $f_\fhalf$\\
\cline{2-4}

$[\Delta\textstyle{{1\over2}}^+]_1(1835)$ & 0.0 $^{+ 1.1}_{- 0.0}$ & 0.0
$^{+ 0.0}_{- 2.2}$ & 0.0 $\pm $ 0.0 & & & & 0.0 $^{+ 2.2}_{- 0.0}$\\%[-9pt]

$\Delta(1740)P_{31}$$^{\rm a}$\\

$[\Delta\textstyle{{1\over2}}^+]_2(1875)$ & 0.1 $^{+ 0.2}_{- 0.1}$ & 0.0
$\pm $ 0.1 & 0.0 $\pm $ 0.0 & & & & 0.1 $^{+ 0.2}_{- 0.1}$\\%[-9pt]

$\Delta(1910)P_{31}$****\\

& $p_\half$ & $p_\thalf$ & $f_\fhalf$\\
\cline{2-4}

$[\Delta\textstyle{{3\over2}}^+]_3(1915)$ & 0.0 $\pm $ 0.0 & - 0.1 $^{+
0.1}_{- 0.3}$ & 0.0 $\pm $ 0.0 & - 0.1 $^{+ 0.1}_{- 0.2}$ & 0.0 $\pm $ 0.0 & &
0.1 $^{+ 0.3}_{- 0.1}$\\%[-9pt]

$\Delta(1920)P_{33}$***\\

$[\Delta\textstyle{{3\over2}}^+]_4(1985)$ & 0.8 $^{+ 3.8}_{- 0.8}$ & 0.3
$^{+ 1.4}_{- 0.3}$ & 0.1 $^{+ 1.3}_{- 0.1}$ & 0.0 $^{+ 0.0}_{- 0.1}$ & - 0.1
$^{+ 0.1}_{- 0.8} $ & & 0.9 $^{+ 4.3}_{- 0.9}$\\

& $f_\half$ & $p_\thalf$ & $f_\thalf$ & $p_\fhalf$ & $f_\fhalf$ & $h_\fhalf$ \\
\cline{2-7}

$[\Delta\textstyle{{5\over2}}^+]_1(1910)$ & 0.0 $\pm $ 0.0 & - 0.1 $^{+
0.1}_{- 0.3}$ & 0.0 $\pm $ 0.0 & - 0.1 $^{+ 0.1}_{- 0.2}$ & 0.0 $\pm $ 0.0 & 
0.0
$\pm $ 0.0 & 0.1 $^{+ 0.3}_{- 0.1}$\\%[-9pt]

$\Delta(1750)F_{35}$$^{\rm b}$\\%[-9pt]
$\Delta(1905)F_{35}$****\\

$[\Delta\textstyle{{5\over2}}^+]_2(1990)$ & 0.0 $^{+ 0.0}_{- 0.5}$ & 0.8
$^{+ 3.6}_{- 0.8}$ & 0.1 $^{+ 1.5}_{- 0.1}$ & -1.3 $^{+ 1.3}_{- 5.6}$ &
- 0.1 $^{+ 0.1}_{- 2.4} $ & 0.0 $\pm $ 0.0 & 1.5 $^{+ 7.2}_{- 1.5}$\\%[-9pt]

$\Delta(2000)F_{35}$**\\

& $f_\half$ & $f_\thalf$ & $h_\thalf$ & $p_\fhalf$ & $f_\fhalf$ & $h_\fhalf$ \\
\cline{2-7}

$[\Delta\textstyle{{7\over2}}^+]_1(1940)]$ & 0.0 $\pm $ 0.0 & 0.0 $\pm $
0.0 & 0.0 $\pm $ 0.0 & -1.0 $\pm $ 0.2 & 0.0 $\pm $ 0.0 & 0.0 $\pm $ 0.0 & 1.0
$\pm $ 0.3\\%[-9pt]

$\Delta(1950)F_{37}$****\\
\tableline
\noalign{a\ \ First $P_{31}$ state found in Ref.~\cite{MANSA}.}
\noalign{b\ \ Ref.~\cite{MANSA} finds two $F_{35}$ states; this one and 
$\Delta(1905)F_{35}$.}
\end{tabular}
\end{table}
\begin{table}

\caption{Results in the $\Delta\omega$ channel for the lightest few
negative-parity $\Delta$ resonances of each $J$ in the N=3 band, and
for the lightest few $\Delta$ resonances for $J^P$ values which first
appear in the N=4, 5 and 6 bands. Notation as in
Table~\protect{\ref{DNle2Deta}}.}

\label{DNge3Domega}
\begin{tabular}{@{}l@{\hspace{6pt}}r@{\hspace{6pt}}r@{\hspace{6pt}}r@{\hspace{6
pt}}r@{\hspace{6pt}}r@{\hspace{6pt}}r@{\hspace{6pt}}r@{}}
\multicolumn{1}{c} {model state}
& \multicolumn{1}{r}{$\Delta\omega$}
& \multicolumn{1}{r}{$\Delta\omega$}
& \multicolumn{1}{r}{$\Delta\omega$}
& \multicolumn{1}{r}{$\Delta\omega$}
& \multicolumn{1}{r}{$\Delta\omega$}
& \multicolumn{1}{r}{$\Delta\omega$}
& \multicolumn{1}{r}{$\sqrt{\Gamma^{\rm tot}_{\Delta\omega}}$}
\\ 
\multicolumn{1}{c} {$N\pi$ state/rating} & & & & & & &\\ 
\tableline
& $s_\half$ & $d_\thalf$ & $d_\fhalf$\\
\cline{2-4}

$[\Delta\half^-]_2(2035)$ & - 0.2 $^{+ 0.2}_{- 0.6}$ & 0.0 $^{+ 0.0}_{- 0.2}$
& 0.0 $\pm $ 0.0 & & & & 0.2 $^{+ 0.7}_{- 0.2}$\\%[-9pt]

$\Delta\half^-(1900)S_{31}$***\\

$[\Delta\half^-]_3(2140)$ & -2.1 $\pm $ 0.7 & 0.1 $\pm $ 0.0 & 5.7 $^{+
5.8}_{- 4.9}$ & & & & 6.1 $^{+ 5.8}_{- 4.4}$\\%[-9pt]

$\Delta\half^-(2150)S_{31}$*\\

& $d_\half$ & $d_\thalf$ & $d_\fhalf$ & $g_\fhalf$\\
\cline{2-5}

$[\Delta\thalf^-]_2(2080)$ & 0.1 $^{+ 1.4}_{- 0.1}$ & - 0.1 $^{+ 0.1}_{-
2.1}$ & 0.0 $\pm $ 0.0 & 0.0 $\pm $ 0.0 & & & 0.1 $^{+ 2.5}_{- 0.1}$\\%[-9pt]

$\Delta\thalf^-(1940)D_{33}$*\\

$[\Delta\thalf^-]_3(2145)$ & 6 $\pm $ 0.6 & 0.2 $\pm $ 0.3 & -4.2 $^{+
3.7}_{- 4.5}$ & 0.0 $\pm $ 0.0 & & & 4.2 $^{+ 4.5}_{- 3.7}$\\

& $d_\half$ & $d_\thalf$ & $g_\thalf$ & $s_\fhalf$ & $d_\fhalf$ & $g_\fhalf$\\
\cline{2-7}

$[\Delta\fhalf^-]_1(2155)$ & 0.0 $\pm $ 0.1 & 0.0 $\pm $ 0.1 & 0.0 $\pm $
0.0 & -1.0 $^{+ 0.7}_{- 1.4}$ & - 0.1 $\pm $ 0.0 & 0.0 $\pm $ 0.0 & 1.0 $^{+
1.4}_{- 0.7}$\\%[-9pt]

$\Delta\fhalf^-(1930)D_{35}$***\\

$[\Delta\fhalf^-]_2(2165)$ & -1.0 $^{+ 0.9}_{- 1.0}$ & -1.0 $^{+ 0.9}_{-
1.0}$ & - 0.5 $^{+ 0.5}_{- 1.5}$ & -3.3 $\pm $ 0.8 & -1.8 $^{+ 1.5}_{-
1.6}$ & 0.0 $\pm $ 0.0 & 4.0 $^{+ 2.3}_{- 1.5}$\\

$[\Delta\fhalf^-]_3(2265)$ & 1.7 $\pm $ 0.5 & - 0.7 $\pm $ 0.2 & 0.3 $\pm $
0.2 & - 0.5 $^{+ 0.1}_{- 0.3}$ & -3.0 $^{+ 0.9}_{- 0.7}$ & -2.4 $^{+ 1.3}_{ -
1.9}$ & 4.3 $^{+ 1.9}_{- 1.5}$\\

$[\Delta\fhalf^-]_4(2325)$ & - 0.2 $\pm $ 0.1 & -1.5 $\pm $ 0.7 & - 0.2 $^{+
 0.2}_{- 0.3}$ & 1.0 $^{+ 0.7}_{- 0.2}$ & 0.2 $\pm $ 0.1 & -1.1 $^{+ 0.8}_{ -
 1.6}$ & 2.1 $^{+ 1.8}_{- 1.0}$\\

& $g_\half$ & $d_\thalf$ & $g_\thalf$ & $d_\fhalf$ & $g_\fhalf$ & $i_\fhalf$\\
\cline{2-7}

$[\Delta\shalf^-]_1(2230)$ & 0.1 $^{+ 0.2}_{- 0.1}$ & - 0.5 $\pm $ 0.3 & 0.3
 $^{+ 0.6}_{- 0.3}$ & -4.1 $\pm $ 2.2 & -1.4 $^{+ 1.1}_{- 2.4}$ & 0.0
 $\pm $ 0.0 & 4.4 $^{+ 3.0}_{- 2.4}$\\

$[\Delta\shalf^-]_2(2295)$ & 0.1 $^{+ 0.2}_{- 0.1}$ & - 0.5 $\pm $ 0.3 & 0.3
 $^{+ 0.6}_{- 0.3}$ & -4.1 $\pm $ 2.2 & -1.4 $^{+ 1.1}_{- 2.4}$ & 0.0
 $\pm $ 0.0 & 4.4 $^{+ 3.0}_{- 2.4}$\\

& $g_\half$ & $g_\thalf$ & $i_\thalf$ & $d_\fhalf$ & $g_\fhalf$ & $i_\fhalf$\\
\cline{2-7}

$[\Delta\nhalf^-]_1(2295)$ & 0.9 $^{+ 1.1}_{- 0.7}$ & 0.5 $^{+ 0.6}_{-
0.4}$ & 0.0 $\pm $ 0.0 & -9.5 $^{+ 4.9}_{- 1.4}$ & -1.9 $^{+ 1.5}_{-
2.2}$ & 0.0 $\pm $ 0.0 & 9.8 $^{+ 2.2}_{- 5.1}$\\%[-9pt]

$\Delta\nhalf^-(2400)G_{39}$**\\

& $f_\half$ & $f_\thalf$ & $h_\thalf$ & $p_\fhalf$ & $f_\fhalf$ & $h_\fhalf$ \\
\cline{2-7}

$[\Delta\shalf^+]_2(2370)$ & 1.4 $\pm $ 0.7 & 0.8 $\pm $ 0.4 & 0.0 $\pm $
0.0 & -3.0 $^{+ 0.2}_{- 1.4}$ & -3.1 $^{+ 1.7}_{- 1.5}$ & 0.0 $\pm $ 0.0 &
4.6 $^{+ 2.1}_{- 1.4}$\\%[-9pt]
 
$\Delta\shalf^+(2390)F_{37}$*\\

$[\Delta\shalf^+]_3(2460)$ & 0.1 $\pm $ 0.0 & -1.0 $\pm $ 0.6 & - 0.2 $^{+
 0.2}_{- 0.3}$ & 0.3 $\pm $ 0.0 & -1.6 $^{+ 0.9}_{- 0.8}$ & -1.1 $^{+ 0.9}_{
 - 1.4}$ & 2.3 $^{+ 1.6}_{- 1.3}$\\

& $h_\half$ & $f_\thalf$ & $h_\thalf$ & $f_\fhalf$ & $h_\fhalf$ & $j_\fhalf$ \\
\cline{2-7}

$[\Delta\nhalf^+]_1(2420)$ & 0.2 $^{+ 0.4}_{- 0.1}$ & -1.3 $^{+ 0.8}_{-
1.2}$ & - 0.1 $^{+ 0.1}_{- 0.2}$ & -1.4 $^{+ 0.9}_{- 1.3}$ & - 0.3 $^{+
0.3}_{- 0.7} $ & 0.0 $\pm $ 0.0 & 1.9 $^{+ 2.0}_{- 1.3}$\\%[-9pt]

$\Delta\nhalf^+(2300)H_{39}$**\\

$[\Delta\nhalf^+]_2(2505)$ & - 0.4 $\pm $ 0.3 & 2.4 $^{+ 0.5}_{- 0.9}$ &
 1.0 $^{+ 0.9}_{- 0.6}$ & -3.1 $^{+ 1.2}_{- 0.7}$ & -1.3 $^{+ 0.8}_{-
 1.1}$ & 0.0 $\pm $ 0.0 & 4.3 $^{+ 1.5}_{- 1.8}$\\

& $h_\half$ & $h_\thalf$ & $j_\thalf$ & $f_\fhalf$ & $h_\fhalf$ & $j_\fhalf$ \\
\cline{2-7}

$[\Delta\lhalf^+]_1(2450)$ & 0.4 $\pm $ 0.3 & 0.2 $\pm $ 0.2 & 0.0 $\pm $
0.0 & -5.1 $^{+ 2.2}_{- 1.6}$ & 0.8 $^{+ 0.6}_{- 0.5}$ & 0.0 $\pm $ 0.0 &
5.2 $^{+ 1.7}_{- 2.3}$\\%[-9pt]

$\Delta\lhalf^+(2420)H_{3\,\,11}$****\\

& $j_\half$ & $h_\thalf$ & $j_\thalf$ & $h_\fhalf$ & $j_\fhalf$ & $l_\fhalf$ \\
\cline{2-7}

$[\Delta\tthalf^+]_1(2880)$ & 0.6 $^{+ 0.6}_{- 0.3}$ & -1.3 $\pm $ 0.5 &
 - 0.3 $^{+ 0.2}_{- 0.3}$ & -1.5 $\pm $ 0.5 & - 0.9 $^{+ 0.5}_{- 0.9}$ & 0.0
 $\pm $ 0.0 & 2.3 $^{+ 1.3}_{- 0.9}$\\

$[\Delta\tthalf^+]_2(2955)$ & - 0.5 $^{+ 0.3}_{- 0.4}$ & 1.6 $\pm $ 0.4 &
1.2 $^{+ 1.0}_{- 0.6}$ & -1.9 $\pm $ 0.6 & -1.4 $^{+ 0.7}_{- 1.1}$ & 0.0
$\pm $ 0.0 & 3.2 $^{+ 1.5}_{- 1.2}$\\

& $i_\half$ & $i_\thalf$ & $k_\thalf$ & $g_\fhalf$ & $i_\fhalf$ & $k_\fhalf$ \\
\cline{2-7}

$[\Delta\tthalf^-]_1(2750)$ & 0.6 $^{+ 0.4}_{- 0.2}$ & 0.3 $\pm $ 0.2 & 0.0
$\pm $ 0.0 & -4.3 $^{+ 0.8}_{- 1.0}$ & -1.1 $^{+ 0.4}_{- 0.6}$ & 0.0 $\pm
$ 0.0 & 4.5 $^{+ 1.2}_{- 0.9}$\\%[-9pt]

$\Delta\tthalf^-(2750)I_{3\,\,13}$**\\

& $j_\half$ & $j_\thalf$ & $l_\thalf$ & $h_\fhalf$ & $j_\fhalf$ & $l_\fhalf$ \\
\cline{2-7}

$[\Delta\fthalf^+]_1(2920)$ & 0.7 $\pm $ 0.3 & 0.3 $\pm $ 0.2 & 0.0 $\pm $
0.0 & -3.6 $\pm $ 0.8 & 1.2 $^{+ 0.6}_{- 0.4}$ & 0.0 $\pm $ 0.0 & 3.9 $\pm
$ 1.0\\%[-9pt]

$\Delta\fthalf^+(2950)K_{3\,\,15}$**\\

$[\Delta\fthalf^+]_2(3085)$ & 0.2 $\pm $ 0.1 & 0.1 $\pm $ 0.1 & 0.0 $\pm $
0.0 & -1.2 $^{+ 0.3}_{- 0.2}$ & - 0.4 $\pm $ 0.2 & 0.0 $\pm $ 0.0 & 1.3 $\pm
$ 0.3\\

\end{tabular}
\end{table}

\end{document}